# Large transverse shifts appearing upon passage of vortices through oblique dark solitons

Yaroslav V. Kartashov[*] and Anatoly M. Kamchatnov

*Institute of Spectroscopy, Russian Academy of Sciences, Troitsk, Moscow Region, 142190, Russia*
*Corresponding author: Yaroslav.Kartashov@icfo.es*

We show that when a single vortex convected by a flowing background collides with an oblique dark soliton in defocusing Kerr-type medium, it passes through the oblique soliton, but this process is accompanied by a considerable vertical shift of the vortex that grows with decrease of the inclination angle of the dark soliton with respect to the direction of the flow and decreases with increase of the flow (or collision) velocity. In the parameter range where the interaction with a vortex does not lead to drastic deformations of the shape of the oblique dark soliton, this shift can be estimated analytically under the assumption that vortex soliton moves along the "stream lines" that exhibit vertical shift in the vicinity of oblique dark soliton. Similar effect was observed upon collision of dark solitons with moving vortex dipoles or rotating pairs of vortices with equal topological charges.

The interactions between solitons – nonlinear self-sustained waves localized in one or several transverse directions or in time due to the exact balance between nonlinearity and diffraction/dispersion effects – was the subject of steady attention since the early days of nonlinear optics. Already in the first experiments it was demonstrated that one-dimensional bright solitons may behave like particles and exhibit considerable transverse spatial or temporal shifts upon collisions [1,2] that are elastic in materials, where soliton evolution is governed by the integrable evolution equations, and inelastic in systems described by the non-integrable equations. In contrast to bright solitons, whose interaction depends on their phase difference, one-dimensional dark solitons in local nonlinear materials usually repel each other [3,4]. In two-dimensional geometries more complex interaction scenarios are possible, when two oblique dark soliton stripes may experience considerable lateral shifts and deformations around their intersection point [5,6].

The interactions between bright or dark solitons of the same dimensionality are now well studied (see reviews [7,8] and references therein). However, much less attention was paid to the interaction between nonlinear excitations of different dimensionalities, especially in defocusing nonlinear media, where in addition to dark solitons, localized in only one transverse direction and infinitely extended in the orthogonal direction, another very interesting type of self-sustained excitation may exist in the form of vortices or vortex dipoles. In optics, such excitations were observed in a number of materials and settings, including photorefractive [9,10] and photovoltaic [11] crystals, and rubidium vapors [12]. Besides that, vortices were found in Bose-Einstein condensates [13-15] and also in polariton condensates incorporating both photon and exciton components [16,17]. It is well-known that two vortices with opposite topological charges may form dipoles moving with the constant velocity in the transverse plane, while equally charged vortices form rotating vortex pairs [18]. Formation of optical vortex dipoles in non-local anisotropic media was studies experimentally in [10]. Such structures naturally appear upon spontaneous decay of dark soliton s [19,20] or they can be generated artificially [21].

Previously the interaction of nonlinear excitations of different dimensionalities (dark solitons and vortices) in defocusing medium was studied only in the case, when collision velocity is zero or when vortex moves in the direction *perpendicular* to the dark soliton stripe. In particular, in [22] the interaction of *static* vortex and dark solitons was considered, when quiescent vortex placed in close proximity of dark soliton leads to its bending and subsequent breakup. In [23] the interaction of moving vortex and dark soliton was treated as a nonlinear analogue of the Aharonov-Bohm scattering. Existence of longitudinal ("horizontal") shift of a vortex upon its passage through the soliton was interpreted as a result of the process of creation of a vortex pair from the dark soliton and subsequent interaction of this pair with the initial vortex. Overall transverse vortex shift was absent here since the velocity of the flow was directed perpendicularly to the dark soliton. Notice that interactions of vortices with edge dislocations in linear medium were studied too [24], although in this case the outcome of interaction is determined by simple interference.

In this paper we address the problem of interaction of a vortex or vortex dipole with an *oblique* dark soliton when the existence of the background flow along the soliton, absent in previous studies [22-24], plays a crucial role in the dynamics of solitons, and predict that the passage of vortex through oblique dark soliton may be accompanied by a considerable shift of the phase dislocation in the direction *transverse* to the direction of the vortex convective motion. This shift can exceed the width of dark solitons by more than one order of magnitude and it is especially pronounced when the angle between the direction of vortex motion and dark soliton is small. We show that in the interaction regime, where collision with a vortex does not lead to strong deformation of an oblique dark soliton, the transverse shift can be calculated analytically under the assumption that vortex soliton is convected along a stream line determined by the initial position of the vortex. It should be stressed that the effect predicted here can be observed in various op-tical media with defocusing nonlinearity, such as photorefractive crystals, liquids with nonlocal defocusing nonlinearity, as well as in Bose-Einstein [25] and polariton [26] condensates, where oblique dark soliton can be generated in the supersonic flow past repulsive defects and where they can become effectively stable for large flow velocities.

We consider the propagation of light beams along the $\xi$ axis of the medium with the defocusing cubic nonlinearity that can be described by the nonlinear Schrödinger equation for the dimensionless amplitude of the light field $q$:

$$i\frac{\partial q}{\partial \xi}=-\frac{1}{2}\left(\frac{\partial^2 q}{\partial \eta^2}+\frac{\partial^2 q}{\partial \zeta^2}\right)+q|q|^2. \quad (1)$$

Here $\xi$ is the propagation distance, normalized to the diffraction length $k_0 r_0^2$, where $k_0 = 2\pi n_0/\lambda$ is the wavenumber; $\mathbf{r}=\{\eta,\zeta\}$ is the transverse coordinate, normalized to the characteristic scale $r_0$.

We are interested in two types of solutions of Eq. (1): vortex and oblique dark soliton. Stationary vortex solutions can be found numerically in the form $q(r,\psi,\xi)=w(r)\exp(im\psi-i\xi)$, where $\psi$ is the azimuthal angle in the $(\eta,\zeta)$ plane, $m$ is the topological charge of central phase dislocation, and it is assumed without loss of generality that $|q|\to 1$ as $r\to\infty$ [27,28]. Further we consider only simplest vortex solitons with topological charges $m=\pm 1$ due to their exceptional robustness in defocusing medium. The intensity of the vortex $|q|^2$ vanishes in the point where phase is not defined and it approaches the asymptotic value $|q|^2=1$ far from the phase dislocation. Such a vortex can be nested in the tilted-phase background wave:

$$q(r,\psi,\xi)=w(\eta-v\xi,\zeta)\exp(im\psi-i\xi+iv\eta-iv^2\xi/2) \quad (2)$$

where the parameter $v$ determines the velocity of the "flow" and, consequently, the velocity with which vortex is convected along the $\eta$-axis. Since we aim to study the collision of vortex and an oblique dark solitons at a nonzero velocity, we suppose that oblique dark soliton does not move from its initial position in the transverse plane $(\eta,\zeta)$ upon propagation (i.e. in the coordinate frame attached to the background wave such a soliton should move with the velocity $-v$ along the $\eta$-axis). The shape of such oblique dark soliton can be obtained by an appropriate change of the reference frame from the solution [29] known for the case of a non-tilted background wave and it is given by:

$$q(\eta,\zeta,\xi)=\{\chi\tanh[\chi(\eta\cos\phi-\zeta\sin\phi)]-iv\cos\phi\}\times \\ \exp(-i\xi+iv\eta-iv^2\xi/2) \quad (3)$$

where $\chi=(1-v^2\cos^2\phi)^{1/2}$ is the form-factor (it characterizes soliton's width), and $\phi$ is the inclination angle of soliton with respect to the vertical $\zeta$-axis. The "grayness" of such a soliton, i.e. its minimal intensity $v^2\cos^2\phi$, grows with increase of the flow velocity and decrease of the inclination angle. The soliton transforms into plane wave when $v\to\cos^{-1}\phi$. The theory developed in [20,25,26] predicts that for sufficiently large flow velocities (for $v\geq 1.5$) the absolute instability of soliton solutions (3) is replaced by the convective one, when any localized disturbance arising on the soliton shape does not grow, but instead is convected by the flow along the soliton from the point where it has appeared. As a result, such solitons behave as effectively stable objects in experiment and can be excited, for example, in the flow past "repulsive" defects [30,31].

To analyze the interactions between a single vortex (2) and oblique (3) dark soliton we nest them on the common background wave and solve Eq. (1) numerically. In all cases the vortex was initially shifted by the distance $\delta\eta=-30$ with respect to the point $\eta,\zeta=0$ crossing the dark soliton.

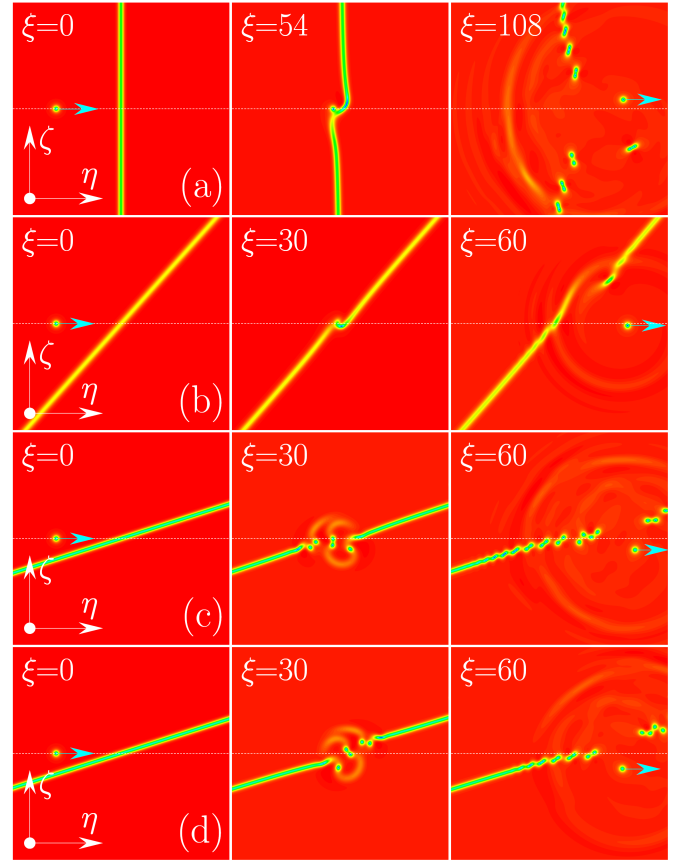

Fig. 1. (Color online) Intensity distributions at different distances illustrating dynamics of interaction of vortex and dark soliton at (a) $v=0.5$, $\phi=0$, $m=-1$, (b) $v=1$, $\phi=0.23\pi$, $m=-1$, (c) $v=1$, $\phi=0.4\pi$, $m=-1$, and (d) $v=1$, $\phi=0.4\pi$, $m=+1$. Horizontal dashed line corresponds to $\zeta=0$. The intensity distributions are shown within the window $\eta,\zeta\in[-50,+50]$. All quantities are plotted in dimensionless units.

Typical interaction regimes are presented in Fig. 1. In general we found two interaction scenarios. First of them takes place when collision velocity is sufficiently small or dark soliton is deep (for $v\geq 1.5$ this condition is met for large inclination angles $\phi$ approaching $\pi/2$, while for $v\ll 1$ even vertical dark stripe can be sufficiently deep). In this case the vortex causes considerable bending of the oblique dark soliton when it approaches the soliton sufficiently close [see Fig. 1(a)], similarly to the interaction of vortices with quiescent solitons [24]. The collision process is accompanied by the formation of vortex dipoles in the region, where dark stripe is strongly curved. Incoming vortex collides with one of such dipoles and fuses with it, replacing the vortex in the dipole with the same topological charge. The vortex that was expelled from the dipole keeps moving along the $\eta$-axis as if it were the initial vortex that has passed through the dark stripe. The collision with slowly moving vortex in this regime results in the rapid development of the snake instability in the oblique soliton [last panel in Fig. 1(a)]. However, when the oblique soliton is relatively shallow and the collision velocity is large, original vortex passes through dark soliton without generation of additional dipoles. Moreover, oblique soliton shows the tendency for restoration after the passage of vortex, since the region where stripe is "broken" is convected into the right upper corner and is gradually replaced by the unperturbed part of soliton [Fig. 1(b)] (see similar behavior at the edges of dark solitons generated by the flow past obstacles in [20]). In any case, in both interaction regimes, at high or low velocities, one clearly observes only one vortex moving in the positive direction of the $\eta$-axis after collision.

The central result of this paper is the appearance of the negative vertical shift [noticeable already in Fig. 1(b)] of the vortex that has passed through oblique soliton. This shift becomes much more pronounced for large inclination angles of the soliton [Fig. 1(c)] and it may considerably exceed its width for sufficiently large $\phi$ values. The overall shift exhibited by the vortex depends also on its topological charge $m$. For selected inclination direction the shift is slightly larger for $m=+1$ vortex than for $m=-1$ one [compare Figs. 1(c) and 1(d)]. This can be understood taking into account the fact that topological charge determines the direction of the velocity around the vortex and, hence, the character of deformation of inclined dark soliton. This deformation appears due to local convection of the soliton segments by the azimuthal velocity created by the vortex [for example, in Fig. 1(a) the stripe is more distorted in the $\zeta<0$ region, while for the opposite vortex charge the deformation would be more pronounced in the $\zeta>0$ region], and it unavoidably affects the trajectory of the vortex.

In the regime where deformations of dark soliton upon interaction are not drastic (i.e., for small inclination angles and large collision velocities) one may calculate the overall shift of the vortex analytically under the assumption that it exactly follows the stream lines that are curvilinear near location of the oblique soliton. Using the approach adopted in hydrodynamics, the field of the dark soliton (3) at $b=-1$ and $\xi=0$ can be written in the form $q=\rho^{1/2}\exp(i\theta)$, where $\rho=|q|^2$ is the intensity of the field and phase is equal to

$$\theta(\eta,\zeta) = v\eta - \arctan\frac{v\cos\phi}{\chi\tanh[\chi(\eta\cos\phi-\zeta\sin\phi)]}. \quad (4)$$

The components of the flow velocity $\mathbf{u}=(u_\eta, u_\zeta)$ are given by the derivatives $u_\eta = \partial\theta/\partial\eta = (\rho\sin^2\phi+\cos^2\phi)v/\rho$ and $u_\zeta = \partial\theta/\partial\zeta = (\rho-1)v/\rho$. The stream lines are determined by the equation $d\eta/u_\eta = d\zeta/u_\zeta$ for a given velocity field $\mathbf{u}$. In particular, the vertical displacement of the stream line $\delta\zeta = \int_{-\infty}^{+\infty}(u_\zeta/u_\eta)d\eta$ upon transition from the $\eta=-\infty$ to $\eta=+\infty$ region (hence, equal to the vortex displacement) is given by

$$\delta\zeta = -\frac{2}{\cos\phi(1+v^2\sin^2\phi)}\arctan\left(\frac{\cos^{-2}\phi-v^2}{\sin^{-2}\phi+v^2}\right)^{1/2}. \quad (5)$$

Analytically predicted dependence $\delta\zeta(\phi)$ for two representative velocity values is shown in Figs. 2(a) and 2(b) with red solid line. As one can see, analytically predicted shift of vortex soliton diverges when $\phi\to\pi/2$ and vanishes at $\phi=0$ for $v\leq 1$ or at $\phi=\arccos(1/v)$ for $v>1$. This curve can be compared with numerically calculated [with the aid of full evolution Eq. (1)] dependencies $\delta\zeta(\phi)$ for both values of topological charge $m$. Notice remarkable agreement of the analytical prediction and numerical results close to the point where oblique soliton are shallow and also for the intermediate values of the inclination angle. The difference between numerical and analytical results is most pronounced at large $\phi$ values where interaction with the vortex results in strong deformation of the dark soliton and, consequently, leads to strong distortion of the velocity field. The initial horizontal position $\delta\eta=-30$ of the vortex was fixed in all simulations, and therefore the initial distance between the vortex and the soliton becomes relatively small for large inclination angles $\phi$. Hence, the interaction between the vortex and the soliton is large enough already at the initial stage of evolution. As a result, the numerically calculated dependence of $\delta\zeta$ on $\phi$ saturates at a certain level, rather than diverges. Notice that in the parameter region, where theory provides relatively accurate value of the transverse shift, it falls exactly between two curves obtained numerically for $m=-1$ and $m=+1$ cases (i.e. $|\delta\zeta|$ is always slightly smaller for negatively charged vortex than for positively charged one). The dependence of shift on the velocity $v$ for a fixed dark soliton inclination angle $\phi=0.35\pi$ is shown in Fig. 2(c). As expected, the best agreement can be observed for large collision velocities, while for small velocities the actual shift strongly exceeds analytical prediction.

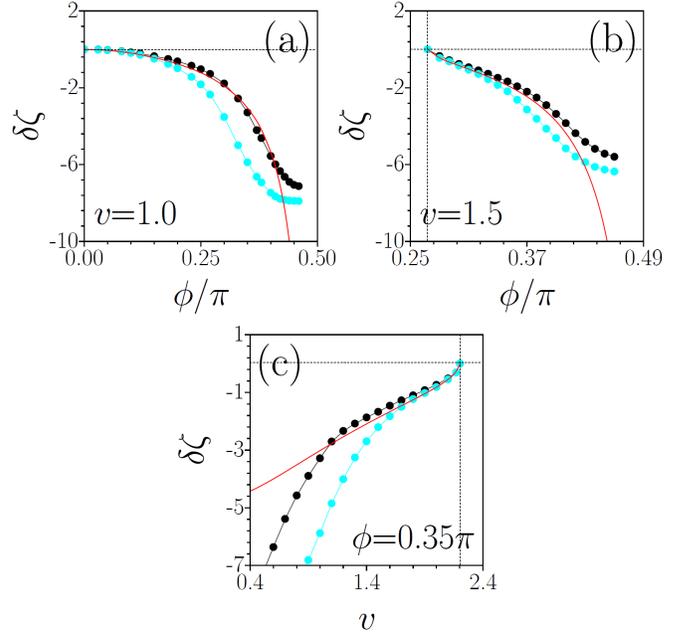

Fig. 2. (Color online) Vertical shift $\delta\zeta$ of vortex soliton passing through dark soliton stripe versus inclination angle $\phi$ of the stripe at $v=1.0$ (a) and $v=1.5$ (b) and versus collision velocity $v$ at $\phi=0.35\pi$ (c). Vertical dashed lines indicate the points where the condition $v=1/\cos\phi$ is met and dark soliton stripe disappears. In all plots solid red line shows analytical prediction, while lines with black (blue) dots show numerical results for vortices with topological charge $m = -1$ $(m = +1)$. Notice that the shift may considerably exceed the width of the vortex core. All quantities are plotted in dimensionless units.

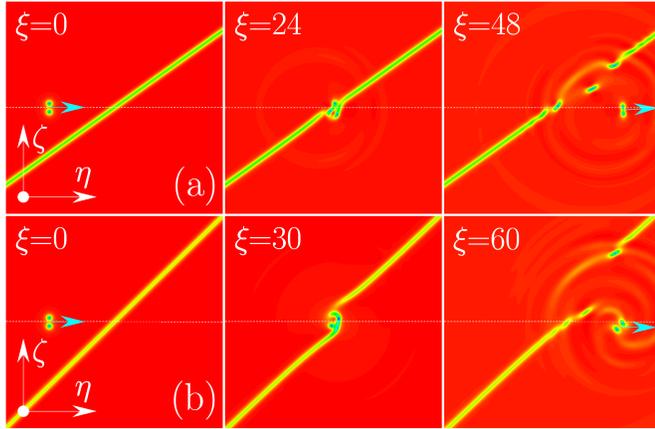

Fig. 3. (Color online) Intensity distributions at different distances illustrating dynamics of interaction of (a) vortex dipole consisting of two vortices with opposite charges $m = \pm 1$ and a dark soliton at $v = 1$, $\phi = 0.3\pi$, and (b) of vortex pair composed of two $m = +1$ vortices and a dark soliton with $v = 1$, $\phi = 0.25\pi$. In both cases the initial separation between vortices in the dipole along the $\zeta$-axis is equal to $4$. Horizontal dashed line corresponds to $\zeta = 0$. The intensity distributions are shown within the window $\eta, \zeta \in [-50, +50]$. All quantities are plotted in dimensionless units.

Finally, it should be stressed that similar effects can be observed upon interaction of oblique dark solitons with more complex vortex states consisting of two oppositely or equally charged vortices. Vortex dipole composed of two spatially separated oppositely charged vortices moves in the direction normal to the line connecting centers of phase dislocations even for zero phase tilt $v$ of the common background [18]. The velocity of its motion $v_d$ increases with decrease of the vertical separation between dislocations. For $v \neq 0$ the actual velocity of vertically oriented vortex dipole is given by $v + v_d$. In addition to vertical shift, vortex dipole can change the direction of its motion after interaction with dark soliton [one example is shown in Fig. 3(a)]. This effect is most pronounced for small collision velocities. In contrast to vortex dipoles, the pair of equally charged vortices exhibit rotation upon propagation. Such a pair perturbs oblique dark soliton even stronger than single vortex does [Fig. 3(b)], but the dependencies of the transverse shift on inclination angle and collision velocity remain qualitatively similar to those presented in Fig. 2.

Summarizing, we showed that the interaction of dark vortices with oblique solitons may be accompanied by a considerable vertical shift of the vortex.